\documentclass[%
 reprint,
 amsmath,amssymb,
 aps,
prb,
]{revtex4-2}

\usepackage{graphicx}
\usepackage{dcolumn}
\usepackage{bm}
\usepackage{adjustbox}
\usepackage{hyperref}

\begin{document}

\preprint{APS/123-QED}

\title{Effect of Electron-Phonon Scattering on the Anomalous Hall Conductivity of Fe$_3$Sn: A Kagome Ferromagnetic Metal}

\author{Achintya Low}

\author{Susanta Ghosh}

\author{Soumya Ghorai}

\author{Setti Thirupathaiah}%
 \email{setti@bose.res.in}
\affiliation{Department of Condensed Matter and Materials Physics, S. N. Bose National Centre for Basic Sciences, JD Block, Salt Lake, Sector III, Kolkata, West Bengal, India, 700106.}%


\begin{abstract}
We report on magnetic and magnetotransport studies of a Kagome ferromagnetic metal,  Fe$_3$Sn. Our studies reveal a large anomalous Hall conductivity ($\sigma_{zx}$) in this system,  mainly contributed by temperature independent intrinsic Hall conductivity ($\sigma^{int}_{zx}$=485$\pm$60 S/cm) and  temperature dependent extrinsic Hall conductivity ($\sigma^{ext}_{zx}$) due to skew-scattering. Although $\sigma^{ext}_{zx}$ value is large and almost equivalent to the intrinsic Hall conductivity at low temperatures, it drastically decreases with increasing temperature,  following the relation $\sigma^{ext}_{zx}=\frac{\sigma_{zx0}^{ext}}{(aT+1)^2}$,  under the influence of electron-phonon scattering. The presence of electron-phonon scattering in this system is also confirmed by the linear dependence of longitudinal electrical resistivity at higher temperatures [$\rho(T)\propto T$]. We further find that Fe$_3$Sn is a soft ferromagnet with an easy-axis of magnetization lying in the $\it{ab}$ plane of the crystal with magnetocrystalline anisotropy energy density as large as 1.02 $\times$ 10$^6$ J/m$^3$.
\end{abstract}

\maketitle


\section{Introduction}

Hall effect due to which the fast-moving charge carriers get deflected transversely under the external magnetic fields in metals and semiconductors~\cite{Hall1879} has found lately many real-life technological applications~\cite{Popovic2003, Ramsden2011}. While an ordinary Hall effect has been noticed in the nonmagnetic metals, an anomalous Hall effect (AHE) was found in the collinear ferromagnetic metals~\cite{Hall1880,Nagaosa2010} and in the non-collinear antiferromagnetic metals~\cite{Nakatsuji2015, Nayak2016}. Moreover, the anomalous Hall effect produces substantially higher Hall resistivity than the ordinary Hall effect, and the field dependent Hall resistivity perfectly scales with magnetization~\citep{Hall1880,Nagaosa2010}. Though the origin of AHE in non-collinear AFM metals is widely understood by the presence of non-zero Berry phase in momentum space, several mechanisms were proposed to understand the AHE in collinear ferromagnets.

Foremost, Karplus and Luttinger (KL) predicted that the AHE in ferromagnetic metals originates from the interband scattering of the charge carriers under spin-orbit coupling~\cite{Karplus1954}, which is recently connected to the Berry curvature of the electronic state of solids~\citep{Manna2018}. Since the KL theory does not take into account the impurity scattering effects and mainly talks on the intrinsic band structure, considered as the intrinsic theory of AHE. Later on, Smit $\it{et~al.,}$  proposed an extrinsic theory of AHE by incorporating the impurity scattering~\cite{Smit1958}. The extrinsic AHE happens by two scattering mechanisms, (i) the skew-scattering: whereby the charge carriers scatter asymmetrically by the localized magnetic impurities~\cite{Smit1958} and (ii) the side-jump: whereby the charge carrier takes a small side jump up on scattering with impurity under spin-orbit coupling~\cite{Berger1970, Lyo1972}. Although many ferromagnetic metals are known to show the AHE~\cite{Nagaosa2010}, the Kagome ferromagnets are quite fascinating systems as they show an intrinsic geometrical frustration, leading to several exotic electronic properties such as the flat-bands near the Fermi level~\citep{Sutherland1986,Lieb1989,Leykam2018,Wu2007}, quantum spin-liquid ground state~\citep{Savary2016, Broholm2020}, Chern insulating state~\citep{Kane2005,Huang2015}, Weyl fermions~\citep{Pal2011,Xu2011, Landsteiner2014}, Dirac fermions~\citep{Novoselov2005,Nomura2007, Pal2011}, and magnetic topological Skyrmions~\citep{Nagaosa2013, Fert2017}, manifesting the anomalous and topological Hall effects.

Thus, the ferromagnetic metal Co$_3$Sn$_2$S$_2$ is found to exhibit giant intrinsic anomalous Hall conductivity ($\sigma_{xy}$ $\approx$ 505 S/cm)~\citep{Liu2018} due to the presence of Weyl nodes near the Fermi level~\citep{Morali2019} and Fe$_3$Sn$_2$ is found to show extremely large anomalous Hall conductivity ($\sigma_{xy}$ $\approx$ 1150 S/cm)~\cite{Ye2018} below 2K and large topological Hall conductivity (-0.875$\mu$$\Omega$ cm) above room temperature~\citep{Hou2017,Li2019}. On the other hand, recently, a few reports on polycrystalline Fe$_3$Sn suggested it to be a ferromagnetic metal in which the Fe atoms form a Kagome network in the $\it{ab}$ plane. Further, it was also shown that Fe$_3$Sn exhibits a large magnetocrystalline anisotropy energy~\citep{Sales2014,Vekilova2019} in addition to the anomalous Nernst effect~\citep{Chen2022}. Although the previous report shows temperature-dependent anomalous Hall conductivity to some extent, a thorough understanding of the anomalous Hall effect in Fe$_3$Sn is still missing, especially the influence of electron-phonon scattering on the AHC.

Fe$_3$Sn belongs to the Ni$_3$Sn-type family of crystal structure with an in-plane Kagome network. Unlike its sister compound Mn$_3$Sn which is a non-collinear antiferromagnet metal, Fe$_3$Sn is a collinear in-plane ($\it{ab}$-plane) ferromagnetic metal. In this paper, we mainly focus on the anomalous Hall effect of Fe$_3$Sn as a function temperature. For this,  we have grown high-quality single crystals of Fe$_3$Sn and performed magnetic and magnetotransport studies.  Our results unravel two important contributions to the total Hall conductivity of Fe$_3$Sn. One of them is the temperature independent intrinsic Hall conductivity originated from the electronic band structure and the other one is the temperature dependent extrinsic Hall conductivity originated from the asymmetric skew-scattering. Most importantly, we observe that the extrinsic skew-scattering Hall conductivity strongly depends on the inelastic electron-phonon scattering rate ($\gamma$), $\sigma^{ext}_{zx}=\frac{\sigma_{zx0}^{ext}}{({\gamma}/{\gamma_0}+1)^2}$. In addition, the linear dependence of longitudinal electrical resistivity confirms the presence of electron-phonon scattering at higher temperatures. We further show that Fe$_3$Sn is a soft ferromagnet with an easy-axis of magnetization lying parallel to the $\it{ab}$ plane. We derive a magnetocrystalline anisotropy energy density as large as 1.02 $\times$ 10$^6$ J/m$^3$.

\section{Experimental Details}

\begin{figure}[t]
\includegraphics[scale=0.50]{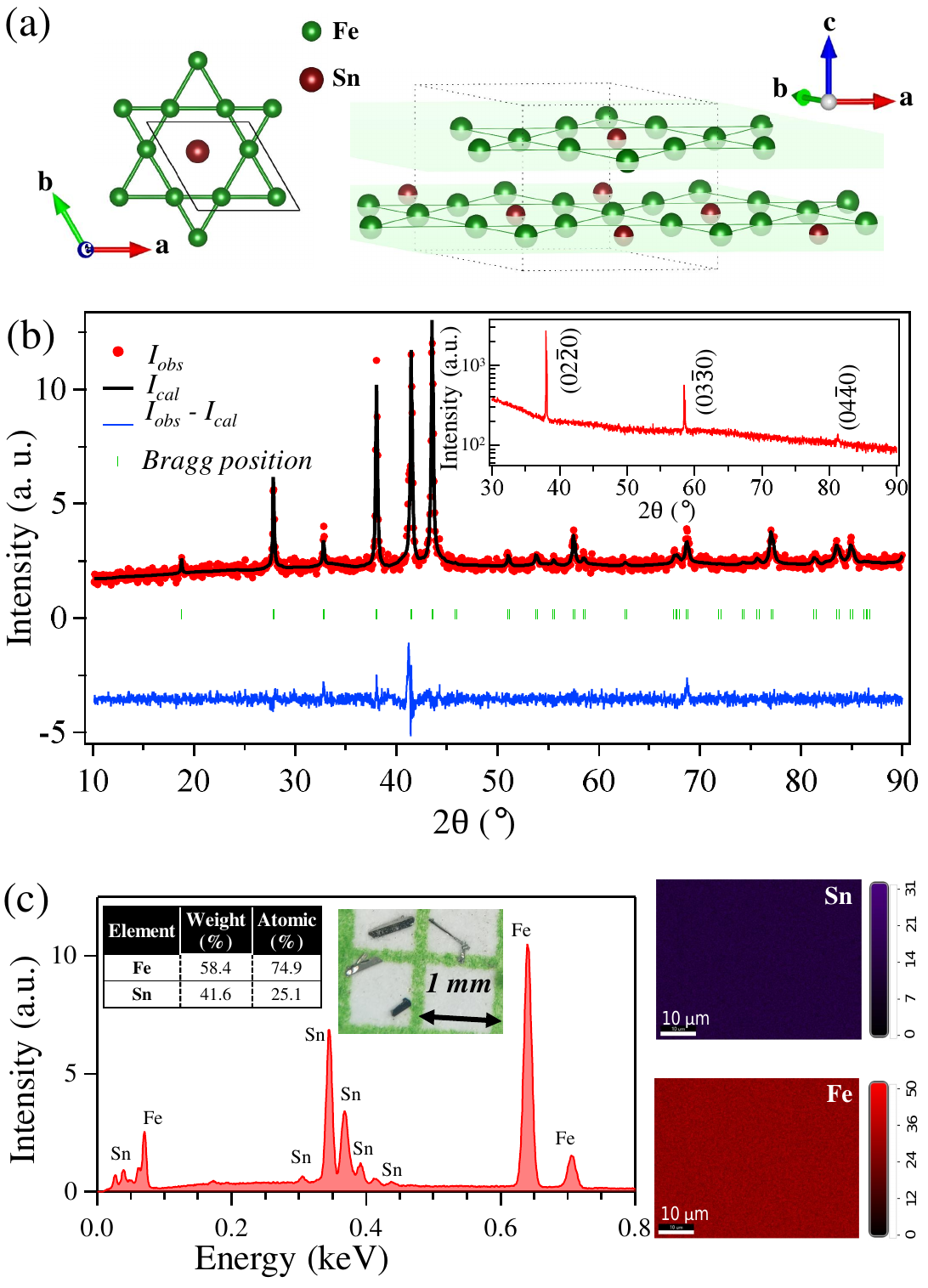}
\caption{(a) Schematic crystal structure and Kagome lattice of Fe$_3$Sn. (b) Powder XRD pattern of crushed Fe$_3$Sn single crystals. Inset of (b) shows intensity reflections correspond to (02$\bar{2}$0) Brag planes. Left panel in (c) presents EDS spectra of Fe$_3$Sn along with tabulated elemental ratios and photographic image showing typical Fe$_3$Sn single crystals.  Right panels in (c) show the elemental mapping of measured single crystal for Fe and Sn.}
\label{Fig1}
\end{figure}

High quality single crystals of Fe$_3$Sn were grown by the solid-state crystal growth (SSCG) technique. In SSCG method,  the crystals are grown out of polycrystalline matrix. Initially, Fe powder (99.99\%, Stern chemicals) and Sn powder (99.995\%, Alfa Aesar) were taken in stoichiometric ratio, grounded thoroughly,  and heated at 810$^o$C for 7 days. As prepared Fe$_3$Sn powder was again grounded and pressed into a pellet which was then annealed at 810$^o$C for another 45 days. Several small rod-shaped shiny crystals with a typical size of 1 mm$\times$0.2 mm$\times$0.2 mm were grown on the surface of the pellet. X-ray diffraction (XRD) technique was performed  on a rod-shaped single crystal and on the crushed crystals using Rigaku SmartLab 9kW Cu K$_\alpha$ X-ray source. Elemental analysis was done using the Energy Dispersive X-ray Spectroscopy (EDS of EDAX) suggests an actual chemical composition of Fe$_{2.98}$Sn,  which is very close to the nominal composition of Fe$_3$Sn.  For the electrical transport and magnetotransport measurements linear four-probe and Hall probe connections were made, respectively,  using the copper wire and silver paint. Magnetic and magnetotransport measurements were carried out on the 9T Physical Properties Measurement Systems (PPMS, Quantum Design-DynaCool) using VSM and ETO options.  To eliminate the longitudinal voltage contribution due to any misalignment of the connections, the Hall resistivity was measured by applying both positive and negative magnetic fields and average Hall resistivity was calculated by $\rho_{H}$=$\frac{\rho_{H}(H)- \rho_{H}(-H)}{2}$.

\section{Results and Discussions}

\begin{figure*}
\includegraphics[width=\linewidth]{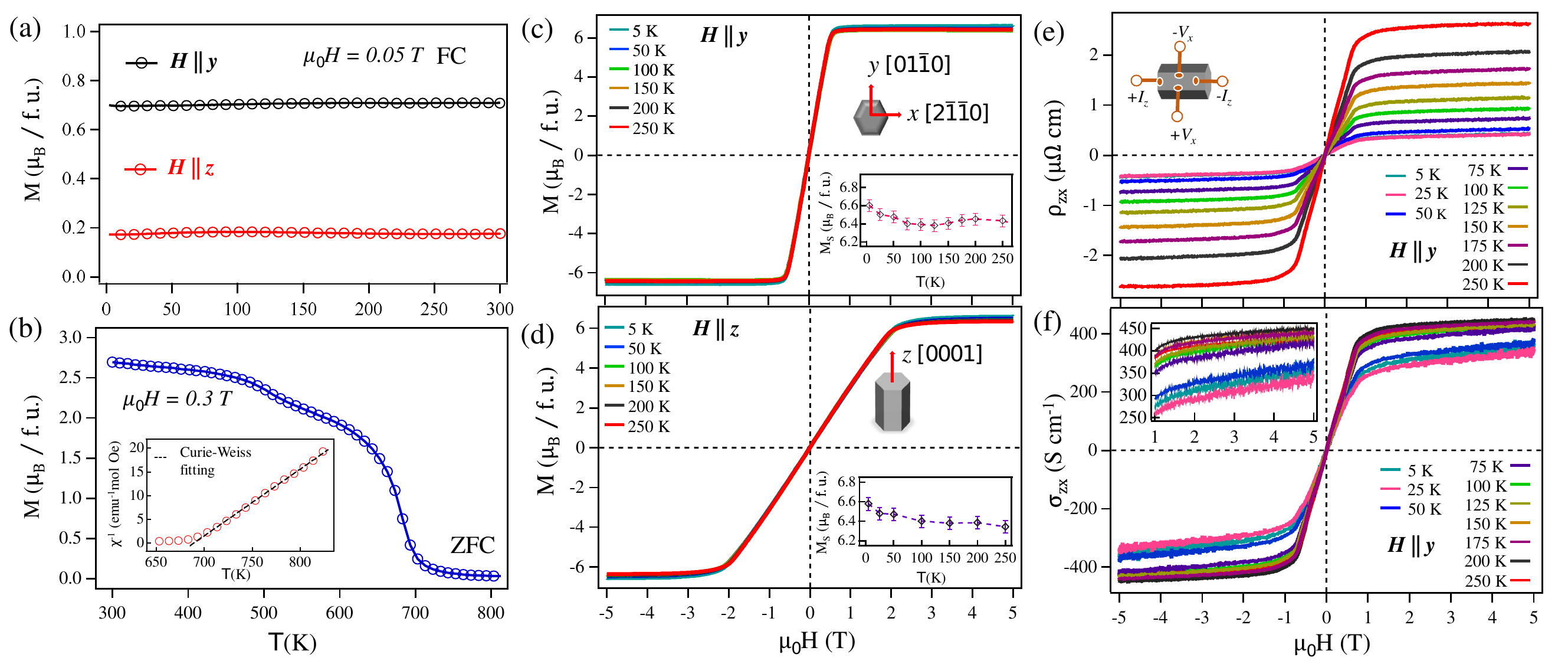}
\caption{(a) and (b) Magnetization measured as a function of temperature on Fe$_3$Sn single crystals and pellets, respectively. Inset of (b) is Curie-Weiss fitting of inverse susceptibility plotted as a function of temperature. (c) and (d) are the magnetization isotherms [$M(H)$] for $H\parallel \it{y}$ and $H\parallel \it{z}$, respectively. Insets in (c) and (d) show the saturation magnetization. (e) and (f) Hall resistivity and Hall conductivity, respectively, plotted as a function of field at various sample temperatures for $H\parallel \it{y}$.}
\label{Fig2}
\end{figure*}

Fig~\ref{Fig1}(a) depicts schematic crystal structure of Fe$_3$Sn where the Fe atoms form a Kagome structure with Sn sitting at the center of the Kagome lattice. Fig.~\ref{Fig1}(b) shows the powder XRD pattern of crushed Fe$_3$Sn single crystals, confirming the hexagonal Ni$_3$Sn type crystal structure with the space group of P6$_3$/mmc (No. 194). Inset of Fig.~\ref{Fig1}(b) depicts the XRD performed on a rod-shaped single crystal, showing intensity of reflections from the ($0~2~\bar{2}~0$) Bragg plane, suggesting that the length of rod-shaped crystals is parallel to the $\it{c}$-axis.  Rietveld refinement performed on the powder XRD of crushed single crystals using the Fullprof software~\cite{RodriguezCarvajal2001}  derives the lattice parameters $\it{a}$=$\it{b}$=5.4631(4)$\AA$ and $\it{c}$=4.3552(4)$\AA$, in good agreement with previous reports~\cite{Sales2014}. Left panel of Fig.~\ref{Fig1}(c) shows the EDS data from which the atomic and weight percentage of the elements are tabulated on the top-left inset of Fig.~\ref{Fig1}(c). Photographic image of typical Fe$_3$Sn single crystals is shown in Fig.~\ref{Fig1}(c).  Elemental mapping performed for Fe and Sn using the EDAX is shown in the right-side panel of Fig.~\ref{Fig1}(c),  implies good homogeneity of single crystals.

Temperature dependent magnetization [$M(T)$] between 2 and 300 K performed on Fe$_3$Sn single crystal with the field ($H$) applied parallel to $\it{y}$-axis ($H\parallel \it{y}$) and parallel to $\it{z}$-axis ($H\parallel \it{z}$) in the field-cooled mode is shown in Fig.~\ref{Fig2}(a). From Fig.~\ref{Fig2}(a) it is evident that below 300 K, $M(T)$ is completely temperature independent. Fig.~\ref{Fig2}(b) shows $M(T)$ performed between 300 and 800 K on Fe$_3$Sn pellet in the zero-field-cooled mode in order to identify the ferromagnetic transition.   From Curie-Weiss fitting of the inverse susceptibility ($\chi^{-1}(T)$), using the formula $\chi=\frac{C}{T-\theta}$, we derive a Curie constant of  $C=7.1\pm 0.1$ $emu.mol^{-1}$Oe$^{-1}$K$^{-1}$ and a Curie-Weiss temperature of $\theta=689\pm2$ K, which are consistent with a previous report~\citep{Sales2014}. The effective magnetic moment of Fe atom in Fe$_3$Sn is found to be $\mu_{eff}$(Fe)=$2.51$ $\mu_{B}$ using the relation $\mu_{eff}$=2.828 $\sqrt{C}$~\cite{Mugiraneza2022}.

Figs.~\ref{Fig2}(c) and ~\ref{Fig2}(d) depict the magnetization isotherms [$M(H)$] measured at various sample temperatures for $H\parallel \it{y}$ and $H\parallel \it{z}$, respectively. The saturation field for $H\parallel \it{y}$ is around 0.5 T whereas it is 2 T for $H\parallel \it{z}$, clearly suggesting that Fe$_3$Sn has an easy-axis of magnetization parallel to the $\it{ab}$-plane.  Also, the absence of hysteresis in the $M(H)$ data for the filed applied parallel both directions makes this system a good soft ferromagnet. The saturation magnetic moment per Fe atom is found to be $M_s$=$2.2$ $\mu_{B}$ which is close to value of effective magnetic moment $\mu_{eff}$=$2.51$ $\mu_{B}$. From the magnetization isotherms shown in Figs.~\ref{Fig2}(c) and ~\ref{Fig2}(d), we estimated the magnetocrystalline anisotropic energy density $K_u$ = $1.02\times 10^6$ J/m$^3$ using the relation $K_u=\mu_0\int_{0}^{M_s}[H_{y}(M)-H_{z}(M)]~dM$ after excluding the geometrical demagnetization factor~\cite{Prozorov2018}. See supplemental material for more details on the calculations of demagnetization factor~\cite{Supple}. Here, $M_s$ represents saturation magnetization, $H_{z}$ and $H_{y}$ represent $H\parallel \it{z}$ and $H\parallel \it{y}$, respectively.

Hall resistivity ($\rho_{zx}$) as a function of field is shown in Fig.~\ref{Fig2}(e) measured at various sample temperatures. Here, the current is applied along the $\it{z}$-axis and the magnetic field is applied along the $\it{y}$-axis to measure the Hall voltage along the $\it{x}$-axis of the crystal as depicted in the inset of Fig.~\ref{Fig2}(e). It is evident from Fig.~\ref{Fig2}(e) that Fe$_3$Sn shows anomalous Hall effect with a little normal Hall effect that is visible at very high temperatures. Next, the Hall conductivity ($\sigma_{zx}$) is calculated using the formula $\sigma_{zx}=-\frac{\rho_{zx}}{{\rho^{2}_{zx}}+{\rho^{2}_{zz}}}$, where $\rho_{zz}$ is the longitudinal resistivity measured along the $\it{z}$-axis of the crystal. Fig.~\ref{Fig2}(f) shows $\sigma_{zx}$  plotted as a function of field in which we observe a large anomalous Hall conductivity in the range of  410-425 S-cm$^{-1}$ between 75 and 250 K when measured at 5 T, and a sudden drop to 370 S-cm$^{-1}$ is observed upon lowering the sample temperature below 75 K. Though high temperature Hall conductivity obtained in this study is consistent with previous report made on polycrystalline Fe$_3$Sn~\cite{Chen2022}, the low temperature Hall conductivity of 200 S-cm$^{-1}$ at 2 K observed in Ref.~\cite{Chen2022} is significantly smaller than our findings ($\approx$ 370 S-cm$^{-1}$ at 5 K).

\begin{figure*}[ht]
\includegraphics[width=\linewidth]{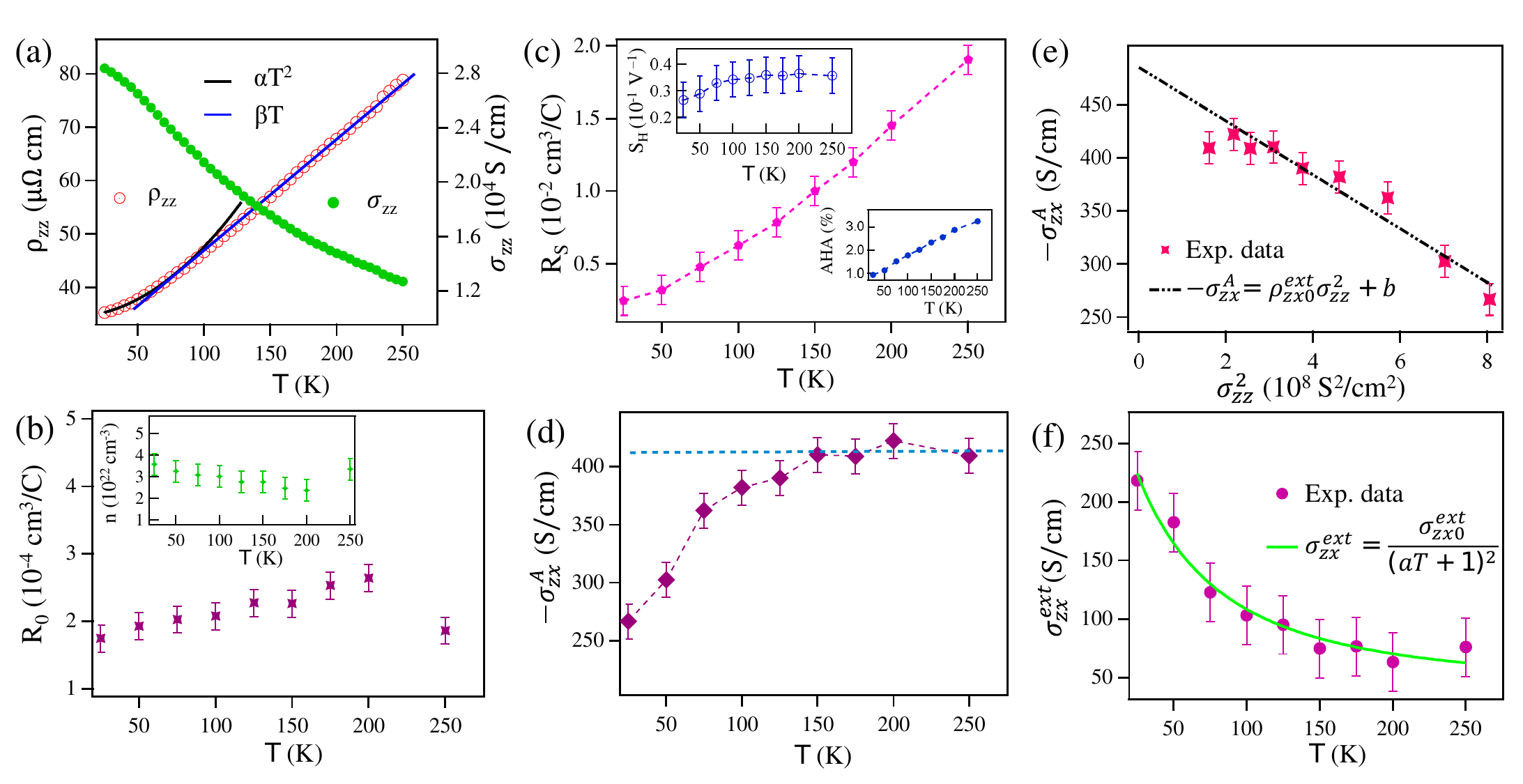}
\caption{(a) Longitudinal electrical resistivity ($\rho_{zz}$) and electrical conductivity ($\sigma_{zz}$) plotted as a function of temperature. (b) and (c) Normal Hall coefficient ($R_0$) and Anomalous Hall coefficient ($R_S$) plotted as a function of temperature, respectively. Charge carrier density ($n$) is shown in the inset of (b).  Top inset in (c) shows anomalous Hall scale factor ($S_H$) and bottom inset in (c) shows anomalous Hall angle percentage (AHA\%). (d) Anomalous Hall conductivity (-$\sigma^{A}_{zx}$) as a function of temperature. (e) -$\sigma^{A}_{zx}$ $vs.$ $\sigma^{2}_{zz}$ plot. The dashed line in (e) is a linear fitting using the relation -$\sigma^{A}_{zx}=\rho^{ext}_{zx0}\sigma_{zz}^2+b$. (f) Extrinsic anomalous Hall conductivity $\sigma^{ext}_{zx}$ plotted as a function of temperature. The solid green curve in (f) is a fit with the equation $\sigma^{ext}_{zx}=\frac{\sigma^{ext}_{zx0}}{(aT+1)^2}$.}
\label{Fig3}
\end{figure*}

Next, Fig.~\ref{Fig3}(a) depicts temperature dependent longitudinal resistivity ($\rho_{zz}$) measured with current applied along the $\it{z}$-axis of the crystal. We notice that $\rho_{zz}$ quadratically depends on the temperature ($\rho_{zz}\propto T^2$) up to $\approx$75 K, demonstrating a Fermi-liquid type nature of the resistivity at low temperatures~\citep{Behnia2022}.  Nevertheless, above 75 K,  $\rho_{zz}$ shows linear depends on the temperature ($\rho_{zz}\propto T$) due to a strong electron-phonon interaction~\citep{Varelogiannis1998}. The temperature dependent resistivity profile of our Fe$_3$Sn single crystal is consistent with previous report on the polycrystalline Fe$_3$Sn~\cite{Chen2022}. Particularly, the linear dependence of $\rho_{zz}$ above 75 K is in very good agreement with Ref.~\cite{Chen2022}. Temperature dependent longitudinal conductivity ($\sigma_{zz}$) also is shown in Fig.~\ref{Fig3}(a). In general, the total Hall resistivity presented in Fig.~\ref{Fig2}(e) can be expressed by the empirical formula,  $\rho_H=\rho^N_H+\rho^A_H$~\citep{Nagaosa2010}. Where, the first term represents normal Hall contribution $\rho^N_H=\mu_0R_0H$ and the second term represents the anomalous Hall contribution which in turn depends on the magnetization ($M$) as $\rho^A_H=\mu_{0}R_SM$. Here, $R_0$ and $R_S$ are normal and  the anomalous Hall coefficients, respectively.  Further, with the help of normal Hall coefficient ($R_0$) one can calculate the charge carrier ($q$) density using the relation, $n=\frac{1}{R_0|q|}$. Since the field dependent anomalous Hall effect is a replica of magnetization, anomalous Hall resistivity saturates beyond a critical field and becomes almost field independent while the normal Hall resistivity linearly depends with field.

Thus, in order to separate the anomalous Hall resistivity from the normal Hall contribution, we fitted the high field region of the total Hall resistivity with a linear function of field and subtracted the normal Hall contribution from the total Hall resistivity. In this way,  we extracted various important parameters such as the normal ($R_0$) and anomalous ($R_S$) Hall coefficients. Fig.~\ref{Fig3}(b) depicts  $R_0$ plotted as a function of temperature. The positive $R_0$ values throughout the measured temperature range as observed in Fig.~\ref{Fig3}(b) suggest hole-carrier dominant electrical transport in Fe$_3$Sn. Inset of Fig.~\ref{Fig3}(b) demonstrates almost temperature independent hole carrier density ($n_h$) which is of the order of $\sim$ 10$^{22}$ $cm^{-3}$, suggesting Fe$_3$Sn to be a good metal. Fig.~\ref{Fig3}(c) shows the anomalous Hall coefficient (R$_S$) plotted a function of temperature. Top-left inset of Fig.~\ref{Fig3}(c) presents the anomalous Hall scaling factor ($S_H$), defined as $S_H=\frac{-\sigma^{A}_{zx}}{M}=\frac{\mu_0 R_S}{{\rho_{zz}}^2}(\cong\frac{{\rho^{A}_{zx}}}{{M\rho_{zz}}^2}$), plotted as a function of temperature. We observe that $S_H$ is almost temperature independent within the error-bars. Moreover, the value of $S_H$=0.03$\pm$0.01 $V^{-1}$ derived in this study is within the range of 0.01-0.2 $V^{-1}$, for any typical ferromagnetic metal~\cite{Zeng2006,Wang2016}. Bottom-right inset of Fig.~\ref{Fig3}(c) shows temperature dependent anomalous Hall angle (AHA), defined as the deviation of electron flow from the current direction, is calculated using the formula AHA(\%)$=\frac{\sigma_{zx}}{\sigma_{zz}}\times 100$(\%).  We clearly notice that AHA(\%)$\approx$3\% at 250 K which decreases with temperature. The value of AHA(\%)$\approx$3\% is close to the AHA values reported on a similar Kagome ferromagnetic system Fe$_3$Sn$_2$  ($\approx$1.1\%)~\cite{Wang2016},  Kagome antiferromagnetic systems such as  Mn$_3$Sn ($\approx$3.2\%)~\cite{Nakatsuji2015} and Mn$_3$Ge ($\approx$5\%)~\cite{Nayak2016}, but much smaller than the Shandite Kagome ferromagnet Co$_3$Sn$_2$S$_2$ ($\approx$20\%)~\cite{Liu2018}.

Several mechanisms were proposed for understanding the anomalous Hall effect in magnetic and nonmagnetic metals. In most of the proposals, the anomalous Hall resistivity ($\rho^{A}_{zx}$) is represented mainly by the function of longitudinal resistivity ($\rho_{zz}$), $\rho^{A}_{zx}=f(\rho_{zz})$.  More explicitly, (i) The intrinsic Karplus-Luttinger mechanism of AHE describes the Hall resistivity $\propto$ $\rho_{zz}^2$  due to the interband scattering in presence of strong spin-orbit coupling~\citep{Karplus1954}, (ii) The extrinsic side-jump mechanism of AHE  describes the Hall resistivity $\propto$ $\rho^{2}_{zz}$  due to side-jump scattering of charge carriers at the impurities~\citep{Berger1970}, (iii) The extrinsic skew-scattering mechanism of AHE describes the Hall resistivity $\propto$ $\rho_{zz}$ due to skew-scattering of charge carrier at the impurities~\citep{Smit1958}, (iv) a $\rho_{zz}^{\alpha}$ ($1<\alpha<2$) dependency of $\rho^{A}_{zx}$ was proposed in the case of bad metals~\citep{Lavine1961}. For our case, the mechanism (iv) can be ignored as the longitudinal conductivity of Fe$_3$Sn is found to be $\sigma_{zz}$=31.4 $\frac{e^2}{h\it{c}}$ (where $\it{c}$ is the lattice constant) which is in the metallic regime~\cite{Shitade2012}.   Recently, a new mechanism (TYJ) was proposed by Tian $et.~ al.$~\citep{Tian2009} in order to understand the AHE in ferromagnetic metals. According to TYJ theory, the anomalous Hall resistivity is described by $\rho^{A}_{zx}=f(\rho_{zz0},\rho_{zz})$ which includes the residual resistivity ($\rho_{zz0}$). Thus, following the TYJ theory~\citep{Tian2009,Hou2015},  the anomalous Hall resistivity takes the form $\rho^A_{zx}=(\alpha \rho_{zz0}+\beta \rho_{zz0}^2)+\it{b}\rho_{zz}^2$ and the anomalous Hall conductivity is represented by $-\sigma^{A}_{zx}=(\alpha \sigma_{zz0}^{-1}+\beta \sigma_{zz0}^{-2})\sigma_{zz}^2+b=\rho_{zx0}^{ext}\sigma_{zz}^2+\it{b}$.  Here, $\alpha$ and $\beta$ are real constants, $\sigma_{zz0}=1/\rho_{zz0}$ is the residual conductivity, and $\it{b}$ is the intrinsic Hall conductivity originated from the momentum space Berry curvature. It is well known that the intrinsic Hall conductivity ($\it{b}$) is usually temperature independent~\citep{Nagaosa2010,Liu2018} except for the systems showing electronic or magnetic phase transitions~\cite{Manna2018,Sung2018}. Since Fe$_3$Sn shows neither electronic nor magnetic transition within the measured temperature range of 2-250 K, change in the Berry phase is not expected. From Fig.~\ref{Fig3}(d), we notice that the anomalous Hall conductivity is almost constant at higher temperatures, but decreases below 150 K. To understand this phenomenon, we employed TYJ mechanism to fit the data of -$\sigma^{A}_{zx}$ $\it{vs.}$ $\sigma^{2}_{zz}$ as shown in Fig.~\ref{Fig3}(e), demonstrating a good fitting within the error-bars.  From the fitting , we extracted intrinsic Hall conductivity $\it{b}$=485$\pm$ 60 S/cm that is close to the previously reported values on polycrystalline Fe$_3$Sn ($\approx$ 500 S/cm)~\citep{Chen2022} and predicted by a theoretical calculation ($\approx$ 600 S/cm at $E_F$)~\cite{Shen2022}.

Next, Fig.~\ref{Fig3}(f) shows the extrinsic Hall conductivity ($\sigma^{ext}_{zx}$) extracted from the total conductivity by subtracting the intrinsic Hall contribution ($\it{b}$). From Fig.~\ref{Fig3}(f), it is evident that the extrinsic Hall conductivity decreases with increasing sample temperature. Further, the sign of extrinsic Hall conductivity is opposite to the intrinsic Hall conductivity, resulting into the reduced total Hall conductivity at lower temperatures. As we know, in the clean limit, for $\hbar/\tau \rightarrow 0$ ($\tau$ is the relaxation time) the extrinsic skew-scattering contribution diverges~\citep{Onoda2006}. On the other hand, from Fig.~\ref{Fig3}(f),  we can see that as the sample temperature decreases the extrinsic Hall conductivity rapidly increases. This observation indicates that skew-scattering playing a major role to generate the anomalous Hall conductivity in Fe$_3$Sn. The longitudinal resistivity ($\rho_{zz}$) suggests electron-phonon scattering at higher temperatures in this system. In order to understand the phonon influence on the anomalous  Hall conductivity, we employed the mechanism proposed by Shitade and Nagaosa~\cite{Shitade2012} which involves the electron-phonon inelastic scattering rates ($\gamma$). In this mechanism, the extrinsic Hall conductivity decays with $\gamma$ following the relation $\sigma_{zx}^{ext}=\frac{\sigma_{zx0}^{ext}}{(\gamma/\gamma_0+1)^2}$. Since the inelastic scattering rate is proportional to the longitudinal resistivity [$\gamma\sim(\rho_{zz}-\rho_{zz0}$)] and $\rho_{zz}-\rho_{zz0}$ linearly depends on temperature [($\rho_{zz}-\rho_{zz0})\propto T$] above 75 K [see Fig.~\ref{Fig3}(a)], we can rewrite the equation as $\sigma_{zx}^{ext}=\frac{\sigma_{zx0}^{ext}}{(aT+1)^2}$. Here, $a=0.011\pm0.003$ K$^{-1}$ represents the measure of inelastic electron-phonon scattering strength. In this way, we could fit the  $\sigma_{zx}^{ext}$ data very well as shown in Fig.~\ref{Fig3}(f). Thus, our results clearly demonstrate the influence of electron-phonon interaction on the extrinsic skew-scattering anomalous Hall conductivity which decreases with increasing temperature.  Finally, we would like to mention here that during the course of our manuscript preparation a preprint on the magnetic studies of Fe$_3$Sn single crystals has appeared~\cite{Prodan2023}. The magnetic properties presented in their study are consistent with our findings. Particularly, the magnetocrystalline anisotropy energy density of 1.23 $\times$ $10^{6}$ $J/m^{3}$ reported in Ref.~\cite{Prodan2023} is in good agreement with the value of 1.02 $\times$ 10$^6$ J/m$^3$ found in this study.


\section{Summary}

In summary, we have successfully grown high quality single crystals of Fe$_3$Sn. In this study, we mainly focussed on understanding the anomalous Hall effect as a function temperature. Our results unravel two main contributions to the total Hall conductivity of Fe$_3$Sn. One of them is the temperature independent intrinsic Hall conductivity originated from the electronic band structure and the other one is the temperature dependent extrinsic Hall conductivity due to the asymmetric skew-scattering. Most importantly, we find that the extrinsic skew-scattering Hall conductivity is significantly influenced by the electron-phonon scattering at higher temperatures and obeys the relation $\sigma^{ext}_{zx}=\frac{\sigma_{zx0}^{ext}}{(aT+1)^2}$. In addition, the longitudinal electrical resistivity ($\rho_{zz}$) confirm the presence of electron-phonon scattering as the resistivity linearly depends on the temperature. We show that Fe$_3$Sn is a soft ferromagnet with easy-axis magnetization lying parallel to the $\it{ab}$ plane. We estimate a magnetocrystalline anisotropic energy density as large as 1.02 $\times$ 10$^6$ J/m$^3$ in Fe$_3$Sn.

\begin{acknowledgments}

The authors thank the Science and Engineering Research Board (SERB), Department of Science and Technology (DST), India for the financial support (Grant No. SRG/2020/000393). This research has made use of the Technical Research Centre (TRC) Instrument Facilities of S. N. Bose National Centre for Basic Sciences, established under the TRC project of Department of Science and Technology, Govt. of India.

\end{acknowledgments}

\nocite{*}

\bibliography{F3S.bib}

\section{Supplementary Information}
\pagebreak
\begin{figure*}[ht]
\includegraphics[width=\linewidth]{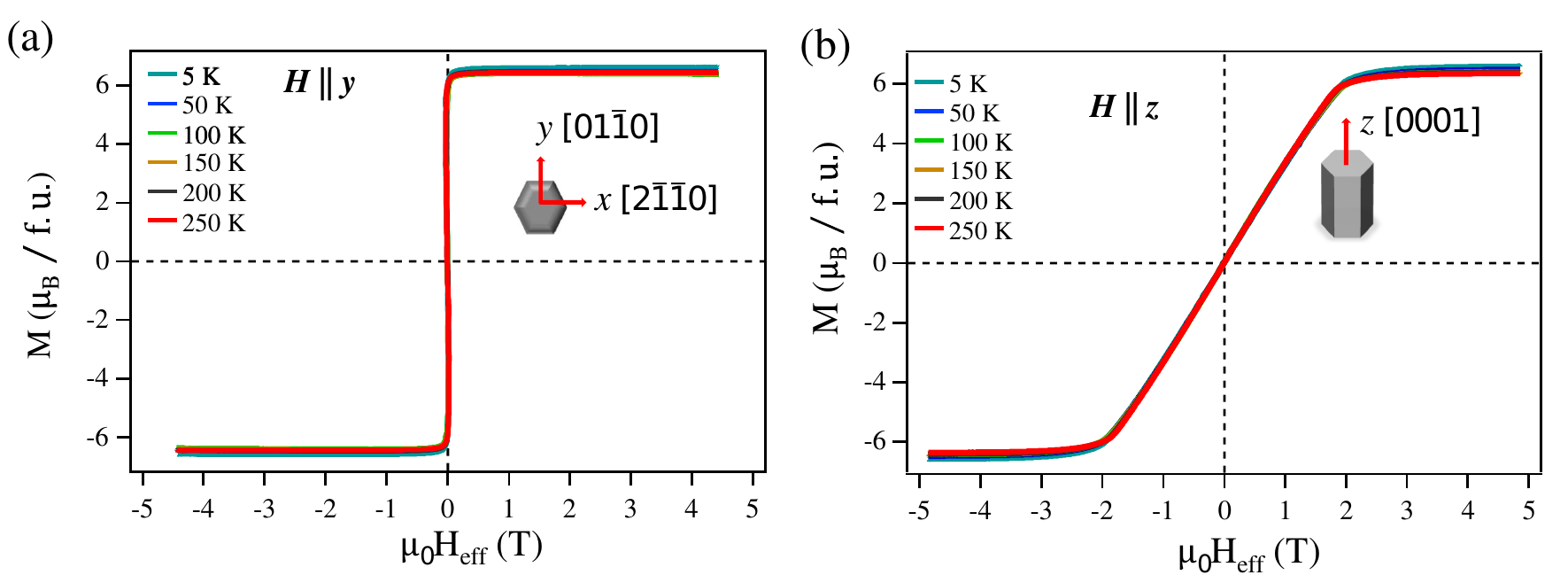}
\caption{Magnetization isotherms [$M(H_{eff})$] measured at various sample temperatures for $H\parallel \it{y}$ (a) and $H\parallel \it{z}$ (b). Here, $H_{eff}$ is the effective magnetization field after excluding the shape demagnetizing factor of the approximately rod-shaped sample using the formula, $H_{eff}=H-N_dM$, where $N_d$ is the demagnetization factor. For our rod-shaped sample, the calculated $N_d^{-1}=2+\frac{1}{\sqrt(2)}\frac{a}{c}$ for $H\parallel\it{y}$ is $0.45$ and $N_d^{-1}=1+1.6\frac{c}{a}$ for $H\parallel\it{z}$ is $0.15$~\cite{Prozorov2018}. Here, $a$=0.2 mm and $c$=0.7 mm are the sample diameter and length, respectively.}
\label{S1}
\end{figure*}

\end{document}